\documentstyle[12pt]{article}

\catcode`@=11 \@addtoreset{equation}{section} \catcode`@=12

\def\strutdepth{\dp\strutbox}
\def\marginarrow#1{\vtop to\strutdepth{
    \baselineskip\strutdepth\vss\llap{#1$\rightarrow$ }\null}}
\def\revision#1{\strut\vadjust{\kern-\strutdepth\marginarrow{#1}}}

\begin{document}
\begin{titlepage}
\begin{flushright}\vbox{\begin{tabular}{c}
           TIFR/TH/97-36\\
           July, 1997\\
           hep-ph/9707315\\
\end{tabular}}\end{flushright}
\begin{center}
   {\large \bf
      Single Polarisation Asymmetries\\
      for Quarkonia in Non-relativistic QCD}
\end{center}
\bigskip
\begin{center}
   {Sourendu Gupta\footnote{E-mail: sgupta@theory.tifr.res.in}\\
    Theory Group, Tata Institute of Fundamental Research,\\
    Homi Bhabha Road, Bombay 400005, India.}
\end{center}
\bigskip
\begin{abstract}
We find that single spin asymmetries in NRQCD are non-vanishing in
general. They are proportional to the imaginary parts of some
non-perturbative matrix elements. With statistics of about $10^6$
identified $J/\psi$'s, or $10^5$ identified $\chi_2$, it is possible
to measure these imaginary parts even if they are an order of magnitude
smaller than the real parts. Such statistics are quite reasonable at
polarised HERA N and other future experiments.
\end{abstract}
\end{titlepage}

\section{\label{intro}Introduction}

Single polarised initial states can give rise to asymmetries in two ways.
If a final state polarisation is observed, then the difference between the
cross sections when the initial and the final state spins are parallel and
anti-parallel gives an asymmetry. These do not yield information beyond what
comes from double polarised initial states. Recent interest has focussed
on single spin asymmetries in which final state spins are not observed---
\begin{equation}
  S \;=\; {\nabla\sigma\over\sigma}\;=\;
      {\sum_h h\sigma(h)\over\sum_h \sigma(h)},
\label{intro.asym}\end{equation}
where $\nabla\sigma$ is the difference of cross sections with the initial
particle polarised in opposite directions. Such asymmetries do not exist
in leading twist computations in perturbative QCD. They seem to arise either
from next-to-leading twist-3 effects, or through non-perturbative effects
like intrinsic transverse momenta of partons.

A formalism for computing the production cross-sections of heavy quarkonia,
called non-relativistic QCD (NRQCD) \cite{bbl}, has been developed.
In NRQCD the non-perturbative long-distance physics is factored from the
short distance perturbative physics. This seems like an ideal place to
search for single polarised asymmetries of the kind in eq.\ (\ref{intro.asym}).
In this paper we report a set of computations which shows that such
asymmetries indeed exist in NRQCD, yield new and interesting non-perturbative
information about quarkonia, and are measurable.

NRQCD is a low-energy effective theory for quarkonia. The action is written
in terms of all possible operators consistent with the symmetries of QCD.
All momenta in NRQCD are cut off by some scale $\Lambda$. The coupling
associated with each term is identified through a perturbative matching
procedure. 
The applicability of NRQCD to quarkonium production depends on the proof that
final state effects factorise in the hadronisation of a heavy quark pair to a
quarkonium. Such a proof has been given for production at large transverse
momenta \cite{bbl} and successful phenomenology has been done \cite{jpsi}.
For low-energy production of quarkonia a proof is lacking, but a reasonable
phenomenology arises when cross sections are computed assuming factorisation
\cite{short,br,long}. Double polarisation asymmetries have been computed for
quarkonia in this framework \cite{pol,polo,hpol}. Numerical results for single
polarisation asymmetries computed in the colour singlet model have been quoted
in the literature \cite{ansel}, and preliminary results in NRQCD 
obtained in \cite{pol}.

The NRQCD factorisation formula for the inclusive production of heavy
quarkonium resonances $H$ with 4-momentum $P$ can be written as
\begin{eqnarray}
   d\sigma\;=\;&
      {\displaystyle{1\over\Phi}{d^3P\over(2\pi)^3 2E_{\scriptscriptstyle P}}}
     \sum_{ij} C_{ij}
          \left\langle{\cal K}_i\Pi(H){\cal K}^\dagger_j\right\rangle,\\
   d\nabla\sigma\;=\;&
      {\displaystyle{1\over\Phi}{d^3P\over(2\pi)^3 2E_{\scriptscriptstyle P}}}
       \sum_{ij} {\overline C}_{ij}
          \left\langle{\cal K}_i\Pi(H){\cal K}^\dagger_j\right\rangle.
\label{intro.nrqcd}\end{eqnarray}
where $\Phi$ is a flux factor. The coefficient functions $C_{ij}$ and
${\overline C}_{ij}$ are computable in perturbative QCD and hence have an
expansion in the strong coupling $\alpha_{\scriptscriptstyle S}$ (evaluated
at the NRQCD cutoff). Although each matrix element in the sum above is
non-perturbative, it has a fixed scaling dimension in the quark velocity $v$.
Then the NRQCD cross section is a double
series in $\alpha_{\scriptscriptstyle S}$ and $v^2$. For charmonium a numerical
coincidence, $\alpha_{\scriptscriptstyle S}\approx v^2$, makes this expansion
a delicate one to handle. For bottomonium, the series is easier to analyse
since $v^2\ll\alpha_{\scriptscriptstyle S}$.

The fermion bilinear operators ${\cal K}_i$ are built out of heavy quark
fields sandwiching colour and spin matrices and the covariant derivative
${\bf D}$. The composite labels $i$ and $j$ include the colour index $\alpha$,
the spin quantum number $S$, the number of derivative operators $N$, the
orbital angular momentum $L$, the total angular momentum $J$ and the helicity
$J_z$. The hadron projection operator 
\begin{equation}
   \Pi(H)\;=\;{\sum_s} \left|H,s\rangle\langle H,s\right|,
\label{intro.hproj}\end{equation}
(where $s$ denotes hadron states with energy less than the NRQCD cutoff),
is diagonal in $L$, $S$, $J$ and, when the final state spin is unobserved, 
connects states of opposite $J_z$. As a result, the operators ${\cal K}_i$
and ${\cal K}_j$ in eq.\ (\ref{intro.nrqcd}) are restricted to have equal
$L$, $S$, $J$ and opposite $J_z$, for both $d\sigma$ and $d\nabla\sigma$.

The $J_z$-dependence of these matrix elements can be factored out using the
Wigner-Eckart theorem---
\begin{eqnarray}
   \langle{\cal K}_i\Pi(H){\cal K}^\dagger_i\rangle&=&
       {\displaystyle{1\over2J+1}} {\cal O}^H_\alpha({}^{2S+1}L_J^N),
\label{intro.ored}\\
   \langle{\cal K}_i\Pi(H){\cal K}^\dagger_j\rangle&+&{\rm h.c.}\;=\;
       {\displaystyle{1\over2J+1}}
                 {\cal P}^H_\alpha({}^{2S+1}L_J^N,{}^{2S+1}L_J^{N'}).
\label{intro.pred}\end{eqnarray}
The factors of $1/(2J+1)$ come from a Clebsch-Gordan coefficient and are
conventionally included in the coefficient function. Each diagonal matrix
element, ${\cal O}$, is the product of a reduced matrix element
$\langle H||{\cal K}_i^\dagger||0\rangle$ with its complex conjugate
$\langle 0||{\cal K}_i||H\rangle$, and is real. However, the off-diagonal
matrix elements, ${\cal P}$, may have non-vanishing phase. We introduce
the notation
\begin{equation}
   \langle{\cal K}_i\Pi(H){\cal K}^\dagger_j\rangle-{\rm h.c.}\;=\;
    {1\over2J+1}
      {\cal I}^H_\alpha({}^{2S+1}L_J^N,{}^{2S+1}L_J^{N'}),
\label{intro.ired}\end{equation}
where the Wigner-Eckart theorem has been used as before to factor out the
$J_z$ dependence of these matrix elements. The NRQCD power counting rule
for all three types of matrix elements is---
\begin{equation}
   d\;=\;3+N+N'+2(E_d+2M_d),
\label{intro.rule}\end{equation}
where $E_d$ and $M_d$ are the number of colour electric and magnetic
transitions required to connect the hadronic state to the state
${\cal K}_i|0\rangle$. 

In section \ref{te} of this paper we present the main results for
$\nabla\sigma$ after a telegraphic review of the threshold expansion
technique \cite{bchen}, the kinematics and the appropriate Taylor series
expansion of the perturbative matrix element \cite{hiv}. Although the
results presented in this paper are for quarkonia produced at small
transverse momenta, it should be clear that similar effects should arise
also at large transverse momenta. Observable consequences are discussed
in the final section \ref{disc}.

\section{\label{te}Computing the Asymmetry}

We choose to construct the coefficient functions using the ``threshold
expansion'' technique of \cite{bchen}. This consists of calculating, in
perturbative QCD, the matrix element ${\cal M}$ connecting the initial
states to final states with a heavy quark-antiquark pair ($\bar QQ$),
and Taylor expanding the result in the relative momentum of the pair,
$q$, after performing a non-relativistic reduction of the Dirac spinors.
The resulting expression is squared and matched to the NRQCD formula of
eq.\ (\ref{intro.nrqcd}) by inserting a perturbative projector onto a
non-relativistic $\bar QQ$ state between the two spinor bilinears. The
coefficient of this matrix element is the required coefficient function.

In this paper we evaluate the polarised cross sections to order
$\alpha^2_{\scriptscriptstyle S}v^9$. This requires a Taylor expansion
to order, $N+N'\le6$, as can be seen by setting $d=9$ and $E_d=M_d=0$
in eq.\ (\ref{intro.rule}). A simplification occurs because the
perturbative projector has only one term---
\begin{equation}
   \Pi(\bar QQ)\;=\;|\bar QQ\rangle\langle\bar QQ|.
\label{te.qqbproj}\end{equation}
In agreement with \cite{hpol,bchen,hiv} we use the relativistic normalisation
of states
\begin{equation}
   \langle Q(p,\xi)\bar Q(q,\eta) | Q(p',\xi')\bar Q(q',\eta')\rangle
     \;=\;4 E_p E_q (2\pi)^6\delta^3(p-p')\delta^3(q-q'),
\label{te.norm}\end{equation}
with the spinor normalisations $\xi^\dagger\xi=\eta^\dagger\eta=1$. Expanding
$E_p=E_q=\sqrt{m^2+q^2}$ in $q^2$ allows us to write the spinor bilinears in
terms of transition operators built out of the heavy quark field.

The kinematics is very simple to leading order in $\alpha_{\scriptscriptstyle
S}$. The momenta of the initial particles are $p_1$ and $p_2$. We take $p_1$
to correspond to the polarised particle and assume that it lies in the positive
$z$-direction and that $p_2$ is oppositely directed. The net momentum $P=p_1+
p_2$. The 4-momenta of $Q$ and $\bar Q$ ($p$ and $\bar p$ respectively) are
written as
\begin{equation}
   p\;=\;{1\over2}P+L_j q^j\qquad{\rm and}\qquad
   \bar p\;=\;{1\over2}P-L_j q^j.
\label{me.momdef}\end{equation}
Note that $p^2=\bar p^2=m^2$, where $m$ is the mass of the heavy quark.
The space-like vector $q$ is defined in the rest frame of the pair,
and $L^\mu_j$ boosts it to any frame. We shall use Greek indices for Lorentz
tensors and Latin indices for Euclidean 3-tensors. A property of the
boost matrix that we shall use many times is
\begin{equation}
   \epsilon_{\mu\nu\lambda\rho}p_1^\mu p_2^\nu L^\lambda_i L^\rho_j
     \;=\; {M^2\over2}\,\epsilon_{ijk} \hat z_k,
\label{me.rel}\end{equation}
where $\hat z$ is the unit vector in the $z$-direction.
This can be derived from some of the identities listed in \cite{bchen}
and the kinematics given here.

The power counting rule in eq.\ (\ref{intro.rule}) requires that we
express the Fermion bilinear operators as spherical tensors. Recall that
any vector $a_i$ can be written as a spherical tensor of rank 1, with
the components
\begin{equation}
   a_{\pm1}\;=\;\mp{1\over\sqrt2}\left(a_x\pm ia_y\right),\qquad
   a_0\;=\;a_z,
\label{me.spten}\end{equation}
where the subscripts $\pm1$, 0 are helicity indices. In NRQCD
higher rank spherical tensors are constructed by coupling such rank 1
tensors successively \cite{hiv}. Some useful identities are
\begin{eqnarray}
   &a_j b_j\;=\; a_0 b_0 - (a_{+1}b_{-1} + a_{-1}b_{+1})
                                        \;=\;-\sqrt3[a,b]^0_0,
\label{me.stidf}\\
   &i\epsilon_{jkl}a_j b_k \hat z_l\;=\; (a_{+1}b_{-1} - a_{-1}b_{+1}).
\label{me.stide}\end{eqnarray}
We have introduced the notation $[a,b]^J_M$ to denote two spherical tensors
$a$ and $b$ coupled to total rank $J$ and helicity $M$. The coefficient
of the term $[a,b]^0_0$ in eq.\ (\ref{me.stidf}) can be obtained from the
appropriate Clebsch-Gordan coefficients.

\subsection{$\bar qq\to\bar QQ$}

As a simple example of the techniques used, we write down the matrix element
for the subprocess $\bar qq\to\bar QQ$---
\begin{equation}\begin{array}{rl}
   {\cal M}\;&=\;-{\displaystyle{ig^2\over M^2}}
      \left[\bar v(p_2,)\gamma_\mu T^a u(p_1,h)\right] L^\mu_j
     \\ & \qquad\quad \times
    \left[M\xi^\dagger\sigma^jT^a\eta - 
        {\displaystyle{4\over M+2m}}
      \,q^j\xi^\dagger(q\cdot\sigma)T^a\eta\right],
\end{array}\label{me.qqmat}\end{equation}
where $T^a$ is a colour generator, $u$ and $v$ are the light quark spinors
and $\xi$ and $\eta$ are the heavy quark Pauli spinors. The equations of
motion for the initial state quarks have been used to obtain the explicitly
gauge invariant matrix element above. The desired Taylor series expansion
is written down using the relation $M^2=4(m^2+q^2)$ to expand all factors
with $M$. 

The squared matrix element for this process is trivial. The light quark
spinor factors require the projection operator ${p\!\!\!/}_1(1+h\gamma_5)/2$.
The polarised cross section is obtained from the part proportional to $h$.
Doing the Dirac algebra and using the identity in eq.\ (\ref{me.rel}) to
simplify the result, we find---
\begin{equation}
   \nabla|{\cal M}|^2\;=\;-
        {\displaystyle{i\alpha_{\scriptscriptstyle S}^2\over4\pi^2M^4}}
       \epsilon_{jkl}\hat z_l {\bf M}_j{\bf M}_k^\dagger,
\label{me.step1}\end{equation}
where ${\bf M}_j$ is the heavy-quark spinor bilinear in eq.\ (\ref{me.qqmat}).
Using eq.\ (\ref{me.stide}) and the Wigner-Eckart theorem, it is easy to see
that the contribution of the diagonal operators vanishes.

The final results for the cross section differences, to order $v^9$, are---
\begin{equation}\begin{array}{rl}
   \nabla\hat\sigma^{\eta_c}_{\bar qq}\;&=\;
   \nabla\hat\sigma^{h_c}_{\bar qq}\;=\;0,\\

   \nabla\hat\sigma^{J/\psi}_{\bar qq} \;&=\;
       {\displaystyle{\pi^3\alpha_s^2\over54m^6}}\delta(\hat s-4m^2)
          \displaystyle{2\over\sqrt3}
                       {\cal I}^{J/\psi}_8({}^3 S^0_1,{}^3 S^2_1),\\

   \nabla\hat\sigma^{\chi_J}_{\bar qq} \;&=\;
       {\displaystyle{\pi^3\alpha_s^2\over54m^6}}\delta(\hat s-4m^2)
          \biggl[\displaystyle{2\over\sqrt3}
                     {\cal I}^{\chi_J}_8({}^3 S^0_1,{}^3 S^2_1)
                +\displaystyle{7\sqrt5\over12m^2}
                     {\cal I}^{\chi_J}_8({}^3 S^0_1,{}^3 S^4_1)
          \biggr],
\end{array}\label{me.xsecq}\end{equation}
where $\hat s$ is the parton CM energy. The expressions are exactly the same
whether the initial quark or anti-quark is polarised. The matrix elements are
all of order $v^9$. The corresponding unpolarised cross sections have been
written down in \cite{hiv}. They are of lower order in $v$.

The fact that the asymmetry is non-zero for the ${}^3P_0$ state, $\chi_0$,
might come as a surprise. However, one should note that the matrix element
involved in its production entails the radiation of at least one soft gluon.
Thus, in $\bar qq$ annihilation a $\chi_0$ can only be produced along with
some light hadrons. The total angular momentum of the final state is 1, making
it possible to generate this single spin asymmetry.

\subsection{$gg\to\bar QQ$}

The squared matrix element for the $gg$ process is technically a little
more complicated. We work in a class of ghost-free gauges called planar
gauges \cite{ddt}. The density matrix for an initial state gluon of
momentum $p$ and helicity $h$ in these gauges is given by
\begin{equation}
\epsilon_\mu(p,h)\epsilon_\nu^*(p,h)\;=\; {1\over2}
     \left[ -g_{\mu\nu} + {1\over p\cdot V} (p_\mu V_\nu+p_\nu V_\mu)
     + {i h\over p\cdot V} \epsilon_{\mu \nu \rho \sigma} p^\rho V^\sigma 
     \right]. 
\label{me.gauge}\end{equation}
The vector $V$ defines the gauge choice. We write $V=c_1 p_1+c_2 p_2$, with 
$c_1/c_2\sim{\cal O}(1)$.

The sum of the matrix elements arising from the three Feynman diagrams
($s$-channel gluon exchange, ${\cal M}_s$, and $t$- and $u$-channel
quark exchanges, ${\cal M}_t$ and ${\cal M}_u$) can be decomposed into
three colour amplitudes---
\begin{equation}
   {\cal M}\;=\;{1\over6}g^2\delta_{ab}S
               +{1\over2}g^2d_{abc}D^c
               +{i\over2}g^2f_{abc}F^c.
\label{me.ampl}\end{equation}
The colour amplitudes $S$ and $D$ involve only ${\cal M}_t+{\cal M}_u$,
whereas $F$ involves ${\cal M}_s$ as well as ${\cal M}_t-{\cal M}_u$. 

In order to write down our results, we find it convenient to introduce the
notation
\begin{equation}
  {\cal A}\;=\;{1\over M^2}\varepsilon_{\lambda\sigma\mu\nu}
         p_1^\lambda p_2^\sigma\epsilon_1^\mu\epsilon_2^\nu
     \qquad{\rm and}\qquad
  {\cal S}_{ij}\;=\;A_i\hat z_j+A_j\hat z_i-B_{ij}
               +\epsilon_1\cdot\epsilon_2\hat z_i\hat z_j,
\label{me.not4}\end{equation}
where
\begin{equation}\begin{array}{rl}
   A_i\;&=\;{1\over M}\left(\epsilon_1\cdot L_i \epsilon_2\cdot p_1
       -\epsilon_2\cdot L_i \epsilon_1\cdot p_2\right),\\
   B_{ij}\;&=\;\epsilon_1\cdot L_i \epsilon_2\cdot L_j
       +\epsilon_2\cdot L_i \epsilon_1\cdot L_j.
\end{array}\label{me.not2}\end{equation}
Here $\epsilon_i$ is the polarisation vector for the initial gluon of momentum
$p_i$.
In order to identify all terms to order $v^9$ we need the
colour amplitude $S$ to order $q^5$---
\begin{equation}\begin{array}{rl}
   S\;&=\;-\left({\displaystyle8im\over\displaystyle M}\right){\cal A}
                   \,(\xi^\dagger\eta)
       + {\displaystyle 4\over\displaystyle M} {\cal S}_{jm}
                   \,(q^m\xi^\dagger\sigma^j\eta)
       - \left({\displaystyle32im\over\displaystyle M^3}\right){\cal A}
                   \hat z_m\hat z_n\,(q^mq^n\xi^\dagger\eta)
     \\ & \qquad\quad
       + {\displaystyle16\over\displaystyle M^3}
           \left[{\cal S}_{jm}\hat z_n\hat z_p
                     -{\displaystyle M\over\displaystyle M+2m}\delta_{jm}
                {\cal S}_{np}\right]\,(q^mq^nq^p\xi^\dagger\sigma^j\eta)
     \\ & \qquad\qquad
       - \left({\displaystyle128im\over\displaystyle M^5}\right){\cal A}
                   \hat z_m\hat z_n\hat z_p\hat z_r
             \,(q^mq^nq^pq^r\xi^\dagger\eta)
     \\ & \quad\quad
       + {\displaystyle64\over\displaystyle M^5}
           \left[{\cal S}_{jm}\hat z_n\hat z_p
                     -{\displaystyle M\over\displaystyle M+2m}\delta_{jm}
                {\cal S}_{np}\right]\hat z_r\hat z_s
                     \,(q^mq^nq^pq^rq^s\xi^\dagger\sigma^j\eta).
\end{array}\label{me.ampls}\end{equation}
The amplitude $D$ differs only through having colour octet matrix elements
in place of the colour singlet ones shown above. For the colour amplitude
$F$ we need the expansion
\begin{equation}
   F^c\;=\;-\left({\displaystyle16im\over\displaystyle M^2}\right){\cal A}
                   \hat z_m\,(q^m\xi^\dagger T^c\eta)
       + {\displaystyle8\over\displaystyle M^2}{\cal S}_{jm}\hat z_n
                   \,(q^mq^n\xi^\dagger\sigma^jT^c\eta).
\label{me.amplf}\end{equation}
In all three colour amplitudes, the terms in ${\cal A}$ are spin singlet and
those in ${\cal S}$ are spin triplet.

The density matrix in eq.\ (\ref{me.gauge}) yields
\begin{equation}
   \nabla{\cal S}\cdot{\cal S}^*\,\equiv\,
     {1\over4} \sum_h h\, a_j b_m\,S_{jm}S^*_{j'm'}\, c_{j'} d_{m'}\,=\,
    \left\{[a,b]^2_{-2}[c,d]^2_2-[a,b]^2_2[c,d]^2_{-2}\right\},
\label{me.ident}\end{equation}
where $a$, $b$, $c$ and $d$ are Euclidean 3-vectors, written in terms of
the Euclidean components on the left and as spherical tensors on the right.
Gauge invariance is obvious from the fact that the final result does not
depend on the gauge choice $V$. In addition, we find the other gauge
invariant contractions, $\nabla{\cal A}\cdot{\cal A}^*=0$ and
$\nabla{\cal A}\cdot{\cal S}^*\ne0$. Since ${\cal A}$ always comes
with a spin-singlet operator and ${\cal S}$ with a spin-triplet, this
last combination always gives rise to a product of two fermion bilinears
of opposite parity. Such operators vanish in NRQCD. Finally, the single
spin asymmetries can be found by retaining only the terms in ${\cal S}$
in the three colour amplitudes $S$, $D$ and $F$.

After retaining only those spherical tensors which contribute to order
$v^9$ (they are tabulated in \cite{hiv}) and dropping all the diagonal
tensors, very few terms remain. It is clear, for example, that the
colour amplitude $F$ could contribute through only one term. However,
its contribution must then be diagonal, and as a result this colour
amplitude cannot give a single spin asymmetry at this order. Similar
arguments tell us that only two terms contribute from each of the $S$
and $D$ colour amplitudes. The coefficient functions can be read off
from earlier computations of the unpolarised cross section \cite{hiv}.

The direct $J/\psi$ single polarised cross section difference is
\begin{equation}
   \nabla \hat\sigma^{J/\psi}_{gg}(\hat s) \;=\;
          \varphi{5\over48}{\overline\Theta}^{J/\psi}_D(9),
   \qquad{\rm where}\qquad
   \varphi = {\pi^3\alpha_s^2\over4m^2}\delta(\hat s-4 m^2),
\label{me.jpsi}\end{equation}
and ${\overline\Theta}^{J/\psi}_a(d)$ denotes combinations of non-perturbative
matrix elements from the colour amplitude $a$ ($=S$, $D$ or $F$) at order 
$v^d$ for the cross section difference. The matrix element required here is
\begin{equation}
   {\overline\Theta}^{J/\psi}_D(9)\;=\;
       {\displaystyle{2\over\sqrt{15}m^6}}
          {\cal I}^{J/\psi}_8({}^3P_2^1,{}^3P_2^3).
\label{me.jpsime}\end{equation}
The unpolarised cross section \cite{hiv} starts at order $v^7$.
The cross section differences for $\psi'$, $\Upsilon$ and all other ${}^3S_1$
states are given by the same formul\ae; only the appropriate matrix elements
have to be plugged into eq.\ (\ref{me.jpsime}).

Upto order $v^9$, the single polarised subprocess cross section differences
for ${}^1S_0$, ${}^1P_1$ and ${}^3P_{0,1}$ quarkonia vanish---
\begin{equation}
   \nabla \hat\sigma^{\chi_0}_{gg}(\hat s) \;=\;
   \nabla \hat\sigma^{\chi_1}_{gg}(\hat s) \;=\;
   \nabla\hat\sigma^{\eta}_{gg}(\hat s)\;=\;
   \nabla\hat\sigma^{h}_{gg}(\hat s)\;=\;0.
\label{me.zero}\end{equation}
The ${}^3P_2$ cross section difference is
\begin{equation}
   \nabla\hat\sigma^{\chi_2}_{gg}(\hat s)\;=\;\varphi\left({1\over18}\right)
      \left[{\overline\Theta}^{\chi_2}_S(7)+{\overline\Theta}^{\chi_2}_S(9)
                     \right]
\label{me.chi2}\end{equation}
where the combinations of non-perturbative matrix elements are
\begin{equation}
   {\overline\Theta}^{\chi_2}_S(7)\;=\;
      {\displaystyle{2\over\sqrt{15}m^6}}
                {\cal I}^{\chi_2}_1({}^3P_2^1,{}^3P_2^3),
\qquad
   {\overline\Theta}^{\chi_2}_S(9)\;=\;
          {\displaystyle{47\over50m^8}}\sqrt{\displaystyle{3\over7}}
                  {\cal I}^{\chi_2}_1({}^3P_2^1,{}^3P_2^5),
\label{me.chi2me}\end{equation}
The unpolarised cross section \cite{hiv} starts at order $v^5$.

\subsection{$\gamma g\to\bar QQ$ and $\gamma\gamma\to\bar QQ$}

The matrix elements for the two processes $\gamma g\to\bar QQ$ and
$\gamma\gamma\to\bar QQ$ are closely related to the $gg$ amplitudes.
It is obvious that
\begin{equation}
   {\cal M}_{\gamma g}\;=\; ge D,\qquad{\rm and}\qquad
   {\cal M}_{\gamma\gamma}\;=\; e^2 S,
\label{me.gamma}\end{equation}
where $D$ and $S$ are the colour amplitudes given in eq.\ (\ref{me.ampl}),
and $e$ is the charge of the heavy quark.

The $\gamma g$ cross sections for the production of any quarkonium state
can be obtained from those for the $gg$ process, by the prescription---
replace $\alpha_{\scriptscriptstyle S}^2$ by $\alpha\alpha_{\scriptscriptstyle
S}$, delete the ${\overline\Theta}_S$ and ${\overline\Theta}_F$ terms, and
replace the colour factor $5/48$ for the terms in ${\overline\Theta}_D$ by 2.
The $\gamma\gamma$ cross sections are obtained with the prescription---
replace $\alpha_{\scriptscriptstyle S}^2$ in $\varphi$ by $\alpha^2$, delete
the ${\overline\Theta}_D$ and ${\overline\Theta}_F$ terms, and replace the
colour factor $1/18$ for the terms in ${\overline\Theta}_S$ by 16.
Only ${}^3S_1$ quarkonia have non-vanishing single spin asymmetries in
almost-elastic photo-production, and only $\chi_2$ can have non-zero asymmetry
in photon-photon fusion.

\section{Experimental Outlook\label{disc}}

The observable asymmetries in hadro-production are obtained by polarising
one of the initial hadrons. These quantities are obtained by convoluting
the cross sections found in Section \ref{te} with appropriate parton density
functions. It is easy to see that
\begin{equation}
   S^H_{pp}(s,y)\;=\;
      {\nabla\hat\sigma^H_{\bar qq}(\hat s)\nabla{\cal L}_{\bar qq}(s,y)
         +\nabla\hat\sigma^H_{gg}(\hat s)\nabla{\cal L}_{gg}(s,y)\over
       \hat\sigma^H_{\bar qq}(\hat s){\cal L}_{\bar qq}(s,y)
         +\hat\sigma^H_{gg}(\hat s){\cal L}_{gg}(s,y)},
\label{ds.asympp}\end{equation}
where $\sqrt s$ is the centre of mass (CM) energy of the colliding protons
and $y$ is the rapidity at which the final state quarkonium, $H$, is
observed. With the convention that the positive $z$-direction is defined
by the direction of motion of the polarised initial state hadron, the 
parton luminosities can be written as
\begin{equation}
   \nabla{\cal L}_{\bar qq}\;=\;
      \sum_f\Delta q_f(x_1)\bar q_f(x_2)
        +\Delta\bar q_f(x_1)q_f(x_2), \quad
   \nabla{\cal L}_{gg}\;=\;\Delta g(x_1) g(x_2),
\label{ds.lumin}\end{equation}
where $x_1=(2m/\sqrt s)\exp(y)$ and $x_2=(2m/\sqrt s)\exp(-y)$ are the
fractional momenta of the polarised and unpolarised initial hadrons
(respectively) carried by the partons. The quantities $\Delta q_f$,
$\Delta\bar q_f$ and $\Delta g$ are the usual polarised quark, anti-quark
(of flavour $f$) and gluon densities, and the corresponding symbols without
the $\Delta$ stand for the unpolarised densities. The unpolarised parton
luminosities ${\cal L}_{\bar qq}$ and ${\cal L}_{gg}$ are given by similar
expressions where the polarised parton densities are replaced by the
unpolarised densities.

\begin{figure}
\vskip6truecm
\includegraphics{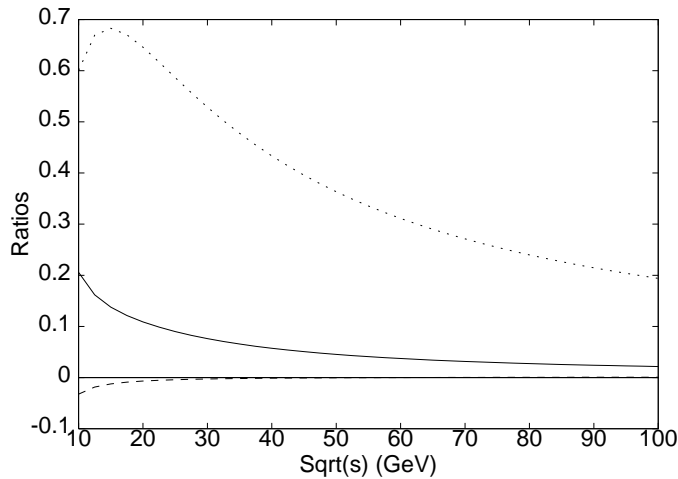}
\caption[dummy]{The ratios ${\cal L}_{\bar qq}/{\cal L}_{gg}$ (full line),
  $\nabla{\cal L}_{\bar qq}/{\cal L}_{gg}$ (dashed line) and
  $\nabla{\cal L}_{gg}/{\cal L}_{gg}$ (dotted line) computed using the GRSV
  LO parton density set \cite{grv} as a function of the CM energy $\sqrt s$
  at zero rapidity.}
\label{fg.lumin}\end{figure}

Numerical values of the ratios of several of these luminosities are plotted
in Figure \ref{fg.lumin} as a function of $\sqrt s$. It is clear from the
figure that ${\cal L}_{gg}$ is the largest and that both the polarised and
unpolarised $\bar qq$ luminosities are small.

The asymmetries $S^{\chi_{c,b}}_{pp}$, and $S^{h_{c,b}}_{pp}$ are likely to
be too small to measure. The ratio of the non-perturbative matrix elements
which determines the analysing power of these processes are of order $v^6$.
Since $v^2\approx0.3$ for charmonia and about $0.1$ for bottomonia, these
asymmetries are unlikely to be observable. For both $S^{\chi_0}_{pp}$ and
$S^{\chi_1}_{pp}$ non-vanishing contributions at order $v^9$ come only from
the $\bar qq$ process whereas the cross sections are dominated by the $gg$
process. Thus, these asymmetries are small because of the smallness of the
ratio $\nabla{\cal L}_{\bar qq}/{\cal L}_{gg}$. $S^{\chi_0}_{pp}$ is further
suppressed by an analysing power which is of order $v^4$.

The most promising observable is the single spin asymmetry for $\chi_2$---
\begin{equation}
   S^{\chi_2}_{pp}(s,y)\;=\;
      \left[{\sqrt{5/3}{\cal I}^{\chi_2}_1({}^3P_2^1,{}^3P_2^3)\over
            m^2{\cal O}^{\chi_2}_1({}^3P_2^1)
           +\sqrt{5/3}{\cal P}^{\chi_2}_1({}^3P_2^1,{}^3P_2^3)}\right]
      \,{\Delta g(x_1)\over g(x_1)}.
\label{ds.schi2}\end{equation}
The analysing power is of order $v^2$; the order $v^4$ correction can be
easily written down using eq.\ (\ref{me.chi2me}), if required. Every matrix
element can be written as a dimensionless number multiplying appropriate
powers of $v$ and the NRQCD cutoff $\Lambda\sim m$. For the real parts of
the matrix elements, there is some evidence \cite{hpol} that this dimensionless
number is independent of which matrix element it comes from, and depends only
on the quarkonium state. We use this assumption along with the definition,
\begin{equation}
   {\cal I}^{\chi_2}_1({}^3P_2^1,{}^3P_2^3)\;=\;
      \tan\Phi_{\chi_2}{\cal P}^{\chi_2}_1({}^3P_2^1,{}^3P_2^3),
\label{ds.phase}\end{equation}
of the phase of this operator\footnote{If the scaling rule in eq.\ 
(\ref{intro.rule}) is to determine the importance of various operators,
then we should expect that the phase is independent of the operator and
depends only on the hadron.} to ask what is the expected sensitivity of
experimental measurements of $\Phi_{\chi_2}$.

First, we rewrite the asymmetry in the form
\begin{equation}
   S^{\chi_2}_{pp}(s,y)\;=\;
      {v^2\sqrt{5/3}\tan\Phi_{\chi_2}\over1+v^2\sqrt{5/3}}
      \,{\Delta g(x_1)\over g(x_1)}.
\label{ds.spr}\end{equation}
Next we note that the measured asymmetry,
\begin{equation}
   S\;=\;{N_+-N_-\over N_++N_-},
\label{ds.trivial}\end{equation}
($N_+$ is the number of events observed when the initial hadron is polarised
in the positive direction, and $N_-$ when it is negative) is smallest and
most susceptible to error when $N_+\approx N_-=N/2$; and the expected error
in $S$ is less than or equal to $1/\sqrt N$ (the equality is reached when the
errors in $N_+$ and $N_-$ are perfectly correlated). By inverting eq.\ 
(\ref{ds.spr}) we can readily see that, near $S=0$, the sensitivity of the
experiment is
\begin{equation}
   \left|\tan\Phi_{\chi_2}\right|\;\ge\;{1+v^2\sqrt{5/3}\over v^2\sqrt{N5/3}}
      \,{g(x_1)\over\Delta g(x_1)}.
\label{ds.spchi}\end{equation}
Recent experiments at $\sqrt s=38.8$ GeV have observed a $J/\psi$ cross
section of about 400 nb in proton nucleon collisions \cite{expts}. About
15\% of these come from decays of $\chi_2$. At the future polarised-HERA N
the CM energy is likely to be about $39.3$ GeV with an integrated
luminosity of 80 ${\rm pb}^{-1}$ per year \cite{ansel}. Even if only the
forward moving $J/\psi$'s are recorded, this should yield about a million
observed $J/\psi$ in the dimuon channel per year of run, and hence about
$10^5$ identified $\chi_2$, assuming a reconstruction efficiency of about
67\% for the $\chi$'s. The appropriate ratio of parton densities\footnote{
By our convention, $x_1$ is the momentum fraction of the parton from the
polarised target, and $x_2$ from the unpolarised beam.} is around
$0.4$. If the degree of polarisation of the target is about 0.8,
polarised-HERA N should be able to measure
\begin{equation}
   \left|\tan\Phi_{\chi_2}\right|\;\ge\;
           0.03\qquad\qquad({\rm single\ polarised-HERA\ N}).
\label{ds.limit}\end{equation}
Thus, single spin asymmetries in $\chi_2$ should be measurable at
polarised-HERA N even if the imaginary part of the matrix element is more
than an order of magnitude smaller than the real part. Current Fermilab
experiments have statistics of about $10^5$ observed $J/\psi$, and hence
about 10000 $\chi_2$ (assuming a reconstruction efficiency of 67\%). With
similar target polarisation their sensitivity is only a factor 3 less.

For direct production of ${}^3S_1$ quarkonia the single spin asymmetry is
given by
\begin{equation}
   S^{\psi'}_{pp}(s,y)\;\approx\;
     \left[{{\overline\Theta}^{\psi'}_D(9)\over
               \Theta^{\psi'}_D(7)+\Theta^{\psi'}_D(9)
                           +(9/5)\Theta^{\psi'}_F(9)}\right]
          \,{\Delta g(x_1)\over g(x_1)},
\label{ds.spsi}\end{equation}
where the unpolarised matrix elements are given in \cite{hiv}. With the
same assumptions as before, we can rewrite this as
\begin{equation}
   S_{pp}(s,y)\;\approx\;
     \left[{0.65v^2\tan\Phi_{\psi'}\over1+1.1v^2}\right]
      \,{\Delta g(x_1)\over g(x_1)}.
\label{ds.sppsi}\end{equation}
The analysis is very similar to that for $\chi_2$. However, the results
are much weaker, since the $\psi'$ cross sections are two orders of
magnitude smaller. In this case the polarised-HERA N and Fermilab
experiments can only measure an imaginary part which is no smaller than
a third of the real part.

The asymmetry for $J/\psi$ is the hardest to predict, due to feed down
from radiative decays of $\chi$ and $\psi'$. The problem can be simplified
by working with a data set consisting of directly produced $J/\psi$. Then
the analysis follows from the analysis of $\psi'$ above. A million $J/\psi$'s
can easily be obtained at polarised-HERA N. The sensitivity is similar to
that for $\chi_2$. At Fermilab fixed target experiments statistics are
poorer by a factor of 10, and the sensitivity is smaller by a factor of 3.

We conclude with a summary of our results. NRQCD predicts non-vanishing
single polarised asymmetries for ${}^3S_1$ and ${}^3P_2$ quarkonia. The
asymmetries are proportional to imaginary parts of some off-diagonal
non-perturbative matrix elements. Since these are unknown, it is not
possible to predict the values of these asymmetries. However,
at polarised-HERA N and with polarised targets at Fermilab it is
possible to measure the values of these imaginary parts even if they
are an order of magnitude smaller than the real parts. Such experiments
are extremely exciting, since they will probe an as yet unknown sector
of the theory of quarkonia.

\end{document}